# Magnetocapacitive $La_{0.6}Sr_{0.4}MnO_3/0.7Pb(Mg_{1/3}Nb_{2/3})O_3 – 0.3PbTiO_3$ epitaxial heterostructures

Ayan Roy Chaudhuri[♣], and S.B. Krupanidhi[♠]
*Materials Research Centre, Indian Institute of Science, Bangalore 560 012, INDIA*

P. Mandal, and A. Sundaresan
*Chemistry and Physics of Materials Unit, Jawaharlal Nehru Centre for Advanced Scientific Research, Jakkur, Bangalore 560064, INDIA*

Abstract:

Epitaxial heterostructures of $La_{0.6}Sr_{0.3}MnO_3$/ $0.7\ Pb(Mg_{1/3}Nb_{2/3})\ O_3 – 0.3\ PbTiO_3$ were fabricated on $LaNiO_3$ coated $LaAlO_3$ (100) substrates by pulsed laser ablation. Ferromagnetic and ferroelectric hysteresis established their biferroic nature. Dielectric behaviour studied under different magnetic fields over a wide range of frequency and temperatures revealed that the capacitance in these heterostructures varies with the applied magnetic field. Appearance of magnetocapacitance and its dependence on magnetic fields, magnetic layer thickness, temperature and frequency indicated a combined contribution of strain mediated magnetoelectric coupling, magnetoresistance of the magnetic layer and Maxwell – Wagner effect on the observed properties.

---

[♣] E-mail: ayan@mrc.iisc.ernet.in
[♠] Corresponding author: S.B. Krupanidhi, E-mail: sbk@mrc.iisc.ernet.in, FAX: +9180 2360 7316.

Multiferroic materials (MF's) have attracted renewed research interest towards understanding the coupling among the electric, magnetic and elastic order parameters due to their enormous scientific interest and significant technological promises[1,2]. Until now, large magnetoelectric (ME) effect has been realized in two major sources: (i) single phase compounds, where presence of multiple long range ordering enhances the internal magnetic and/or electric fields as observed in historical $Cr_2O_3$ [3] and presently focused on Bi based perovskite oxides[4,5], hexagonal RMnO3 (R= Y, In, Ho, Er, Tb, and Lu) manganites, layered manganites e.g. $DyMn_2O_5$, $TbMn_2O_5$ [6] and few other materials, such as $LuFe_2O_4$[7], $CdCr_2SO_4$ [8] etc. (ii) composite materials, laminated structures, and thin film heterostructures consisting of a ferroelectric (FE) and a ferromagnetic(FM) materials, where the ME effect arises as the product property of a magnetostrictive and a piezoelectric compound. Bilayers and multilayers of composites though are especially promising due to their low leakage current and superior poling properties[9-11], often suffer from limitations, such as poor mechanical coupling between layers due to non epitaxial interface, impurities arising from interfacial ion diffusion or reaction under high sintering temperatures, lack of scaling possibilities. Epitaxial multilayered thin films with a coherent and epitaxial interface, can also exhibit negligible ME effect owing to substrate clamping[12]. Therefore selection of lattice matched substrate and materials with very high piezoelectric and magnetostrictive properties, is important to achieve reasonably strong ME coupling. There are reports on dielectric, magnetoresistive, magnetoelectric voltage, spin pinning effect on artificial MF heterostructures consisting of different manganites, such as $La_{1-x}Ca_xMnO_3$, $La_{1-x}Sr_xMnO_3$, etc. as FM material and $BaTiO_3$ (BT), $Ba_{1-x}Sr_xTiO_3$ (BST), $Pb(Zr,Ti)O_3$, $0.7Pb(Mg_{1/3}Nb_{2/3})O_3 – 0.3PbTiO_3$ etc. as the FE material.[13-16]. Until now there is no systematic report on the magnetocapacitive nature of $0.7Pb(Mg_{1/3}Nb_{2/3})O_3 – 0.3PbTiO_3$ (PMNPT)/ $La_{0.6}Sr_{0.4}MnO_3$ (LSMO) bilayered epitaxial thin films. PMNPT having a very high piezoelectric coefficient (~ 1700 pC/N in bulk) compared to other ferroelectric materials like $Pb(Zr, Ti)O_3$, BT, BST might give rise to reasonably strong strain mediated ME coupling with magnetostrictive LSMO. In this letter we have reported the FE, FM, and magnetocapacitive behaviour of PMNPT/LSMO heterostructures fabricated epitaxially on $LaNiO_3$ (LNO) coated $LaAlO_3$ (100) (LAO) substrates.

Thin film PMN-PT/LSMO heterostructures were fabricated by pulsed laser deposition (PLD), using stoichiometric ceramic targets. Details of the growth conditions have been reported elsewhere[17]. The total thicknesses of the bilayered heterostructures were fixed at 240 nm and the individual LSMO layer thickness was varied between 48nm and 120 nm. Crystallographic and epitaxial characterizations of the heterostructures were performed using Phillips X'Pert MRD Pro X Ray diffractometer ($CuK_\alpha$, $\lambda = 0.15418$ nm). The magnetization hysteresis (M-H) was measured using a vibrating sample magnetometer in a PPMS system (Quantum design, USA). A Radiant Technology Precision ferroelectric workstation was used to measure the FE polarization hysteresis (P-E). In order to measure the dielectric response under an applied magnetic field, the samples were mounted on a sample holder inserted in close cycle cryocooled magnet and connected to an Agilent 4294A impedance analyzer using co-axial compensated cables. For all the electrical measurements current perpendicular to the plane geometry has been used with LNO bottom electrode and gold dots of area $3.12\times10^{-3}$ $cm^2$ as top electrode.

Fig.1 presents the X-Ray Diffraction pattern of the PMN-PT/LSMO heterostructures with different individual layer thickness. The sample specifications in this article have been denoted as x/y, where x = thickness of PMNPT in nm and y = thickness of LSMO in nm. Appearance of only (100) and (200) peaks in θ-2θ scan and four fold symmetry observed from Phi scan recorded around the (103) plane of the substrate and the heterostructures for both LSMO and PMN-PT (Inset Fig.1) confirm the "cube on cube" epitaxial growth of the heterostructures. The heterostructures fabricated in this present study are highly strained. Both LSMO (a = 3.86Å) and PMNPT (a = 4.025Å) having respective lattice mismatches of -0.52% and -4.81% with LNO (a= 3.84Å) experience compressive in plane stress. No shift was observed in the XRD peak positions on changing the thickness of individual layers, which indicated that all the heterostructures were under similar strain condition.

M-H hysteresis loops of the 120/120 heterostructure measured at three different temperatures ranging between 20K and 300K exhibited well defined coercivity (Fig. 2), which confirms it's FM nature in the entire range of temperature. Fig. 3(a) shows P-E loops of the 120/120 heterostructure at 20K measured at different probing frequencies ranging between 200 Hz to 2 kHz. FE response of the heterostructure at higher frequency was reflected in the capacitance voltage characteristics at 20K (Fig.3b). The inset A of Fig. 3(a) shows the change of remnant polarization ($\pm P_r$) and coercive voltage ($\pm V_c$) as a function of applied voltage. Both $P_r$ and $V_c$ became gradually flat with applied voltage beyond 15V indicating they approach saturation. The saturated nature of the P-E loops as a function of voltage and their weak frequency dependence are key evidences of the intrinsic FE characteristics of the heterostructures. The total remnant polarization ($2P_r$) and coercive voltage ($V_c$) are 66.7μC/cm$^2$ and 5.96 V respectively at 15 V applied voltage. The 2$P_r$ value is in agreement with the switched polarization value (ΔP) obtained from pulsed polarization positive up negative down (PUND) measurements on the same capacitor with a 15 V pulse of 1ms width and pulse delay 1000ms at 20K. Detailed PUND analysis also exhibited very weak pulse width dependence of ΔP once saturation of P-E loop is reached. Saturated nature of the P-E loops measured at different temperatures (Fig. 3a, inset B) indicates the FE response of the heterostructures over the entire temperature range between 20K and 300K. In the present case the P-E loops also exhibited reduced asymmetry compared to that observed in PMNPT/LSMO superlattices[18], indicating less influence of internal depolarizing field at the polarization and lattice mismatched interface of PMNPT and LSMO. All these observations collectively suggest that the polarization observed is intrinsic to the bilayered heterostructures and does not arise from mobile charges or other extrinsic effects, such as leakage current.

In order to investigate the manifestation of strain mediated magnetoelectric coupling, dielectric responses of the heterostructures have been studied over a wide range of frequency and temperature under different magnetic fields. Since dielectric constant and hence capacitance of FE materials are functions of temperature, at every temperature the system was stabilized before performing the measurements in order to avoid any experimental artifact. Magnetodielectric (MD) effect thus observed is presented by magnetocapacitance (MC), defined as MC(%) = 100× [C(H,T) – C(0,T)]/C(0,T), where C(H,T) represents the capacitance at a magnetic field H and a temperature T, and C(0,T) represents the capacitance without any magnetic field. In analogy to the induced ME voltage, MC might also be proportional to the piezomagnetic coefficient, and hence to the magnetostriction ($\lambda$) of LSMO[9]. Since in plane magnetostriction ($\lambda_{xx}$) of LSMO is at least a factor of two larger than the out of plane magnetostriction ($\lambda_{xz}$)[9], in the present case the magnetic field was applied along the film plane, while the capacitance was measured along the film thickness. Fig.4 shows the MC of the 120/120 heterostructure as a function of frequency at 20K under different magnetic fields ranging between 0.5T – 2T. The MC increased with increasing frequency, took a peak at 260 kHz and decreased again. Under a magnetic field of 2T the peak MC value was 2.3%, corresponding to a change of 40 pF capacitance. MC of the heterostructures was observed to be a function of LSMO layer thickness and increased from 0.5% to 2.3% at the peak frequency with increasing the LSMO layer thickness from 48 nm to 120 nm under same applied magnetic field at 20K. Unlike the ME voltages, which vanishes at higher magnetic fields when $\lambda$ attains saturation[9,19], MC in the heterostructures increased monotonically on increasing the applied magnetic field from 0.2T – 3T in the whole range of frequency. At 5 kHz and 20K, MC and magnetoloss (ML) [ML = 100× {tan$\delta$(H,T) – tan$\delta$(0,T)}/ tan$\delta$(0,T)] vs. magnetic field has been plotted for 120/120 heterostructure in the inset of Fig. 4. Increase in the MC and decrease in ML with increasing magnetic field indicated strong influence of the negative magnetoresistance (MR) of LSMO layer on the observed MD response. MR of LSMO combined with Maxwell – Wagner (MW) effect can show increased positive MC with increasing magnetic field. In such cases MC becomes maximal near the conductivity cutoff frequency in the range of frequency scan under a magnetic field[20]. Manifestation of MR and MW effect was evidenced by the appearance of low frequency (<1 kHz) MC on increasing temperature above 200K when $\lambda_{xx}$ of LSMO decreases[9]. MC effect observed in the present case can thus have combined contribution from both strain mediated ME coupling and MR of the LSMO layer. Optimization of processing parameters and architecture of the heterostructures in order to achieve enhanced strain coupling are currently underway.

In summary, co-existence of FM and FE properties of the epitaxially grown PMN-PT/LSMO heterostructures over a wide range of temperatures established their biferroic nature. MD properties studied over a wide range of temperature and frequency exhibited a clear dependence of capacitance of the heterostructures on the applied magnetic field. Nature of the MC as a function of frequency, applied magnetic field, magnetic layer thickness and temperature collectively suggested a combined contribution from strain coupling and MR effect of LSMO layer on the observed MD properties of the heterostructures.

Figure Captions:

Figure 1. X-Ray diffraction pattern of a PMN-PT/LSMO heterostructures.

Inset: Φ scan of a LAO substrate and a 120nm/120nm heterostructure along PMN-PT (103) plane.

Figure 2. M-H hysteresis loops of PMN-PT/LSMO 120/120 heterostructure at three different temperatures.

Figure 3(a). P-E hysteresis loops of PMN-PT/LSMO 120/120 heterostructure at different probing frequencies at 20K.

Inset A: ±$P_R$ and ±$V_C$ of the heterostructure at different applied voltages.

Inset B: P-E hysteresis loops of PMN-PT/LSMO 120/120 heterostructure at three different temperatures.

Figure 3(b). Capacitance Voltage characteristics of 120/120 heterostructure at 20K.

Figure 4. Magnetocapacitance vs. frequency of 120/120 PMN-PT/LSMO heterostructure at different magnetic fields.

Inset: Magnetocapacitance and magnetoloss of the 120/120 heterostructure as function of magnetic field at 5 kHz.

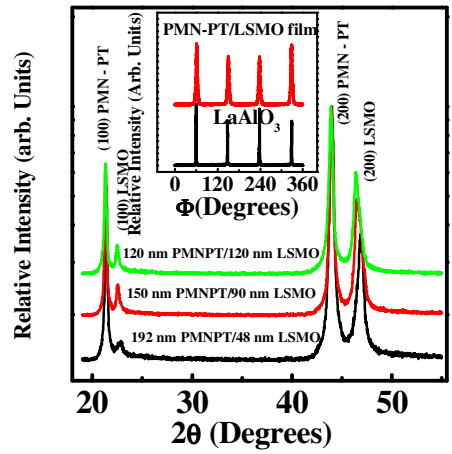

Figure 1

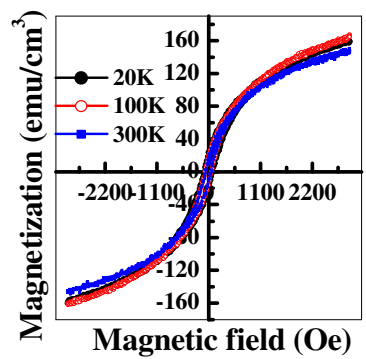

Figure 2

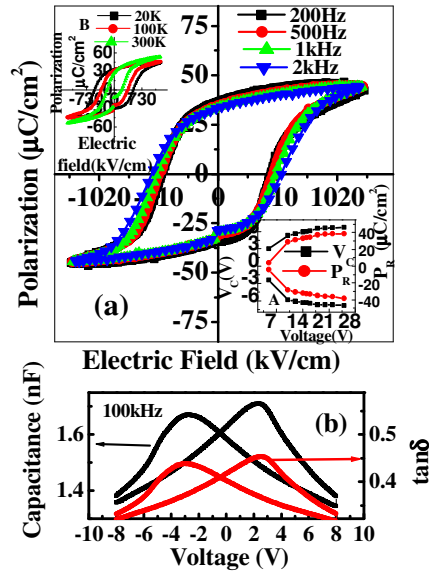

Figure 3

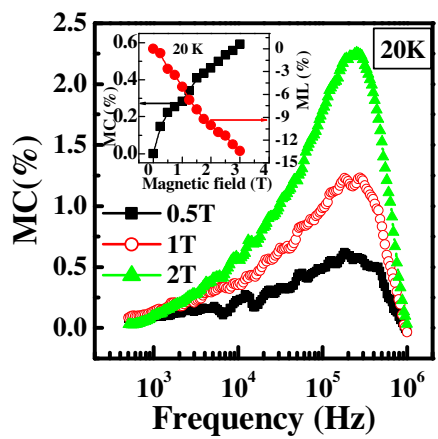

Figure 4